# The Effect of Class Imbalance and Order on Crowdsourced Relevance Judgments

**Rehab K. Qarout, Alessandro Checco, and Gianluca Demartini**
University of Sheffield, Sheffield, UK

## Abstract

In this paper we study the effect on crowd worker efficiency and effectiveness of the dominance of one class in the data they process. We aim at understanding if there is any positive or negative bias in workers seeing many negative examples in the identification of positive labels. To test our hypothesis, we design an experiment where crowd workers are asked to judge the relevance of documents presented in different orders. Our findings indicate that there is a significant improvement in the quality of relevance judgements when presenting relevant results before the non-relevant ones.

## Introduction

Crowdsourcing is often used to scale-out the collection of manual labels that are then used to train machine learning models or to create large Information Retrieval evaluation collections. In such cases, it often happens that the frequency of the classes to be labelled in the data is unbalanced: For example, there are typically few relevant documents as compared to the number of non-relevant ones or there are few fMRI images which are positive to a certain disease as compared to the negative cases.

It is well known that such class imbalance property has negative effects while training supervised machine learning models like, for example, over-fitting the model towards the class which is most frequent in the training data (Ali, Shamsuddin, and Ralescu 2015).

In this paper, we provide preliminary results towards the investigation of the following question: Does class imbalance have effects on the performance of crowd workers involved in the creation of manually labelled datasets?

To answer such question, we run comparative experiments over a popular commercial crowdsourcing platform where we measure label quality and work efficiency over different class distribution settings both including label frequency (i.e., one dominant class) as well as ordering (e.g., positive cases preceding negative ones).

Previous related work has looked at the effect on crowd performances of task granularity (Cheng et al. 2015) and of task type or complexity switch (Lasecki et al. 2015; Cai, Iqbal, and Teevan 2016).

## Experimental Setup

In this study, we used data from the Eighth Text Retrieval Conference (TREC8) (Hawking et al. 2000) which contains a general web search track. From such test collection we took four general topics and for each topic, some documents which have been classified as relevant and some as non-relevant by trained human assessors.

For the current study, we chose one topic with a high number of documents in both the relevant and the non-relevant class, to give more flexibility in the design of the tasks. We chose documents that have similar length and reading difficulty.

To measure effects of class imbalance, we used two different relevant/non-relevant ratios in a batch of judging tasks. The first is composed by 10% relevant and 90% non-relevant documents. The second is composed by 50% relevant and 50% non-relevant documents. For each distribution we created three tasks, each one characterized by a different order of classes in the batch (i.e., relevant first, non-relevant first, and mixed order), as shown in Figure 1. No quality control mechanisms have been implemented. The experiments have been performed on the CrowdFlower platform.

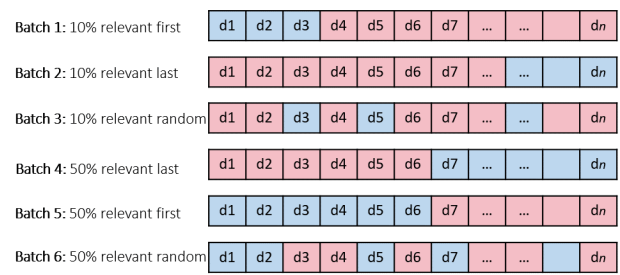

Figure 1: The order of the document classes in each batch, coloured in blue for 'relevant' and red for 'non-relevant'.

We provided the workers with a brief description of the topic and some guide to help them recognise whether a document is relevant to the topic or not. For each batch in Figure 1, we asked 20 workers to judge 30 documents each in one of the pre-defined orders. The reward for completing the entire batch was 4$ per worker. Each worker could only complete one batch to prevent memory bias.

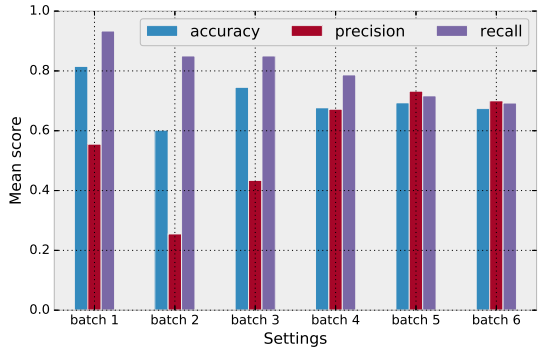

Figure 2: Mean judgement accuracy, precision and recall for each setting.

## Results

In the collected data, we measure worker accuracy in judging the relevance of documents using as ground truth for comparison the original TREC relevance judgements. First, we analyse the judging performance for each of the six batches, as shown in Figure 2. For both distribution types we run a one-way ANOVA test, repeated for each metric (accuracy, precision and recall) with consequent Bonferroni p-value correction of $\alpha = 0.008$. For the first distribution (batches 1, 2 and 3) the precision is statistically significantly different among the three batches ($p = 0.007 < 0.008$). We observe that when the relevant documents are shown before the non-relevant ones (Batch 1) we obtain the highest precision, while the worst precision is obtained when they are shown at the end of the batch (Batch 2). Moreover, in Batch 2 we observe a low number of true positives and a large number of false positive judgements by the workers, which shows how 90% of non-relevant documents shown at the beginning of the batch create a bias in the workers' notion of relevance. Accuracy and recall did not show a statistically significant difference, even if they qualitatively followed the same trend of the precision (maximum for Batch 1 and minimum for Batch 2).

For the second distribution type (50% relevant and 50% non-relevant) there is no statistically significant difference in the performance between different orders: The order of the documents does not influence judgement quality when classes are balanced in the dataset. On the other hand, seeing a similar number of positive and negative documents leads to good performance with more than 60% accuracy in all the three order settings. Overall, best judgement quality in terms of highest accuracy and recall scores has been obtained when relevant documents are presented at the beginning of the sequence.

Finally, we analysed the relation between judgement quality and task completion time, as shown in Figure 3. While no statistically significant trend has emerged among the different settings, for all of them a completion time of less than 500 seconds (to judge 30 documents) has led to an accuracy of less than 0.8. The majority of the workers who spent between 500 and 1800 seconds had an accuracy between 0.6 and 1. We conclude that while introducing lower bounds in task completion time allows to filter out workers who randomly judge relevance, in general completion time is not a sufficient indicator of judgement quality: a result in agreement with (Cai, Iqbal, and Teevan 2016). A more refined analysis of time-based worker profiling, like the one in (Venanzi et al. 2015), is left for future work.

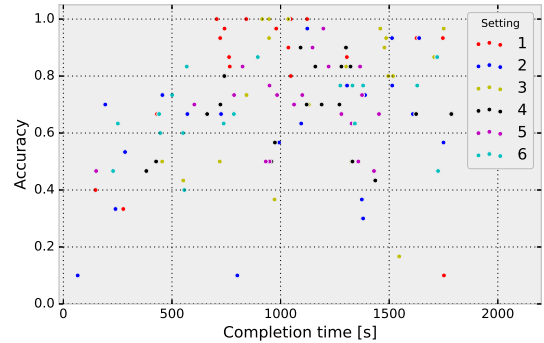

Figure 3: Accuracy vs. completion time for each worker. The colours represent the different batches.

## Conclusions

In this work we looked at the bias effect of class imbalance in a batch of crowdsourced labelling tasks. We specifically focussed on relevance judgement tasks and on how imbalance and order affect worker performances.

We observed that in the case in which most of the documents to be judged are not-relevant and the few relevant ones are presented first, workers perform significantly better.

This is a positive result which can be easily applied in practice as in real IR evaluation settings, e.g., based on pooling documents retrieved by different IR systems, most of the documents to be judged are not-relevant. While in a real setting it is not possible to put relevant documents before non-relevant ones as their relevance label is unknown, it would still be possible to order documents by attributes indicating their relevance (e.g., retrieval rank, number of IR systems retrieving the document, etc.) thus presenting first to the workers the documents with higher probability of being relevant.

In future work we will validate these results with a larger set of experiments including multiple topics and involve a larger number of workers in each experiment. Another possible setting to experiment with in the future is to present few relevant documents first (assumed to be available gold standard data) to prime workers, and then to show a sequence of randomised documents (assumed to have unknown labels).

A different way to deal with the class imbalance problem in crowdsourced labelling tasks is to perform an activity similar to over-sampling for machine learning model training: When a dataset to be labelled is known to be unbalanced, it is possible to introduce additional (possibly artificial) data points to re-balance the dataset. To make such additional tasks useful, they can be used as honey-pot questions with known answers to check for worker reliability.